# Ultrafast Intercavity Nonlinear Couplings between Polaritons

*Jiaxi Wang[1], Bo Xiang[2], Wei Xiong[1,2*]*

[1]*Department of Chemistry and Biochemistry, UC San Diego, La Jolla, CA, 92093*

[2]*Materials Science and Engineering Program, UC San Diego, La Jolla, CA, 92093*

*ORCID: Wei Xiong 0000-0002-7702-0187, Bo Xiang 0000-0002-9055-7931, Jiaxi Wang 0000-0001-7663-6978*

*Abstract* **Realizing nonlinear coupling across space can enable new scientific and technological advances, including ultrafast operation and propagation of information in IR photonic circuitry, remote triggering or catalyzing of chemical reactions, and new platforms for quantum simulations with increased complexities. In this report, we show that ultrafast nonlinear couplings are achieved between polaritons residing in different cavities, in the mid-infrared (IR) regime, e.g. by pumping polaritons in one cavity, the polaritons in the adjacent cavity can be affected. By hybridizing photon and molecular vibrational modes, molecular vibrational polaritons are formed that have the combined characteristic of both photon delocalization and molecular nonlinearity. Thus, although photons have little nonlinear coupling cross-section, and molecular nonlinearity is localized, the dual photon/molecule character of polaritons allows photons to affect each other across different cavities, through coupling to the same molecules – a novel property that neither molecular nor cavity mode would possess alone.**

Achieving delocalized nonlinear interactions is important in advancing many scientific and technology areas, such as ultrafast photonics(1–3), new ways of triggering chemistry remotely(4) and constructing new interacting qubits for quantum simulation and computations(5). Yet, most nonlinearity are local to the sources themselves. For example, molecular modes are a source of optical nonlinearity, which are microscopic and local. In contrast, photon cavity modes can be macroscopic and delocalized, with no nonlinearity. Polaritons, hybrid quasiparticles between molecular and cavity modes(6–30), provide a critical means for achieving delocalized optical nonlinearity by *combining molecular nonlinearity with delocalized cavity modes*(31). For example, a cavity can overlap with neighboring cavities through delocalized evanescent waves, and the molecular modes in the overlapping volumes can enable nonlinear coupling between polaritons in adjacent cavities (Figure 1a), through mechanisms such as excited dark modes(7, 24, 26). We demonstrate this concept here and refer to it as intercavity nonlinear interactions between polaritons.

To realize intercavity nonlinear couplings between polaritons, we fabricate coupled Fabry-Perot (FP) cavity matrices and conduct linear and nonlinear IR spectroscopy on polaritons formed in them. A checkerboard matrix is composed of individual square-shape FP cavities, where cavities with two thicknesses alternate between each other (Figure 1b and c for SEM image). Therefore, cavities with two distinct transition frequencies are aligned next to each other (See Methods). From experiments, we found the optimal cavity lateral dimension to be 50 μm, which ensures a sufficient coupling between neighboring cavities, but avoids the neighbor cavities merging into a single cavity mode due to strong delocalization. The linear IR transmission spectrum shows two sharp cavity transitions at 1970 and 2000 cm$^{-1}$ (Figure 1d), agreeing with expected cavity resonance separations. We prepared the polariton systems by encapsulating a saturated W(CO)$_6$/hexane solution (~40 mM) in the coupled-cavity matrix (see Methods(6, 24)). W(CO)$_6$ has a strong asymmetric stretch vibrational mode at 1983 cm$^{-1}$, which is ideal for forming polaritons in the IR regime.

The linear IR spectra of polaritons in the coupled-cavity matrix show a surprising feature that make it challenging for existing models to capture the results. From a naïve perspective, if the molecular vibrational modes strongly couple to the two cavity modes simultaneously, a system composed of three polaritons, and therefore with three IR peaks, is expected. However, the linear IR spectrum of the coupled-cavity polariton shows a four-peak feature (spectrum on the right part of Figure 1e). This feature indicates that the molecular vibrations might be coupled to the two cavities separately, each forming one pair of polaritons (upper polariton, UP, and lower polariton, LP) to compose the total four peaks, i.e., UP1 and LP1 in cavity A and UP2 and LP2 in cavity B, respectively. To test this idea, we use regular FP cavities to prepare two polariton systems separately, with each one of them matching UP1/LP1 and UP2/LP2, respectively (Figure 1f and g). By simply adding the two spectra together, we can directly compare the combined IR spectrum and the spectrum of polaritons in the coupled cavity. While the polariton peak positions in the combined spectra match well with the ones from the coupled cavity, the intensities do not (Figure 1h). This intensity mismatch suggests that the coupling scenario is not the molecular vibrations coupling to two cavities separately either. In addition, the widely-used semiclassical model(32) (model 1, see S2.1) is unable to reproduce the spectral intensity of the polaritons in the coupled cavity (dashed line in Figure 1h), despite it being able to reproduce the peak position and intensities of polaritons in the regular cavities perfectly (dashed line in Figure 1f and g).

The intensity mismatch suggests a component is missing from existing model to account for the intensity redistribution among polariton peaks. We develop a new model and show the missing component is the delocalization of cavity modes. The essence of model 2 is described in Figure 1b. Upon entering cavity A, after a few round trips, photons can hop to cavity B and subsequently interact with molecular vibrations in cavity B, representing the delocalization between cavities (referred to as the cavity A path). An alternative path also exists (cavity B path in Figure 1b). The expression of transmission spectra based on model 2 is summarized in Eq.1 (detailed derivation in S2.2).

$$T = \left[\frac{T_1 e^{-\frac{1}{2}\alpha L_1}}{1-R_1 e^{i\Delta\varphi_1-\alpha L_1}}\left(1 - R_1^n e^{-n\alpha L_1+in\Delta\varphi_1}\right) + \frac{\sqrt{T_1 T_2} e^{-\frac{1}{2}\alpha L_1}}{1-R_2 e^{i\Delta\varphi_2-\alpha L_2}}\left(R_1^{n-1} e^{-(n-1)\alpha L_1+i(n-1)\Delta\varphi_1} R_2 e^{-\alpha L_2+i\Delta\varphi_2}\right)\right]^2$$

Eq.1

where, $T_m$, $R_m$, $L_m$ and $\varphi_m$ are transmission, reflection, cavity thickness, and phase shift of cavity $m$, respectively, $\alpha$ is the absorptive coefficient of molecules, and $n$ represents the number of round trips before photon hopping to the adjacent cavity. Results from model 2 reproduced not only the spectral peak position but also the intensities from the experimental measurements (Figure 1i). Model 2 indicates that the measured linear IR spectra are a combination of two sets of polaritons from the cavity A and B paths, respectively (orange and cyan traces in Figure 1i). Because of photon hopping, each cavity mode forms three polariton peaks (Figure 1d). The agreement between the simulated and experimental results suggest that cavity mode delocalization do exist among adjacent cavities, a key component for intercavity nonlinear coupling between polaritons (Figure 1a) (31).

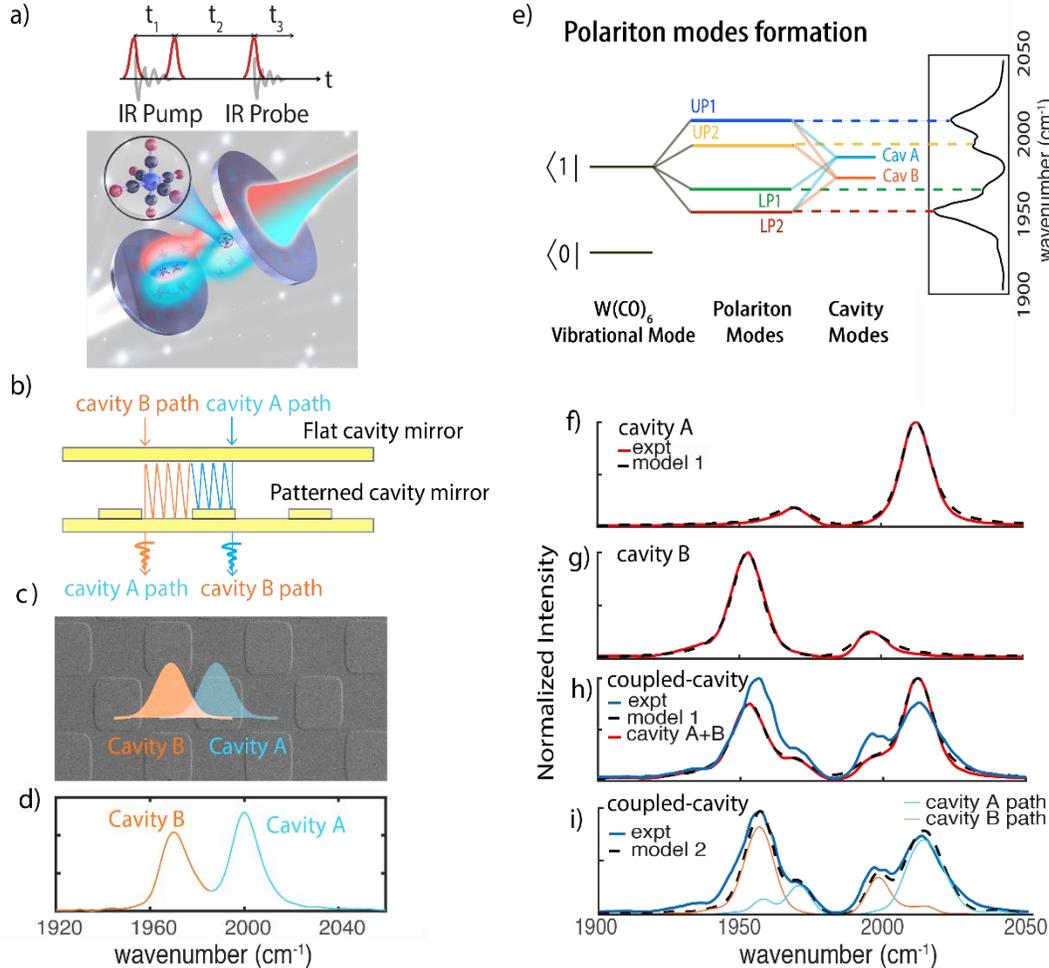

*Figure 1. Key idea of intercavity nonlinear interactions between polaritons. (a) Illustration of a coupled cavity generated using a pair of cavity mirrors and 2D IR pulse sequence. The key to enable intercavity nonlinear interaction is to have anharmonic molecules (enlarged) in the shared volume between cavities. (b) A proposed mechanism of coupling between cavities. When a photon enters into cavity A, it can hop to cavity B and exit from there (cavity A path). The reversed direction (cavity B path) could happen too. (c) SEM of checkerboard-patterned cavity mirror. (d) FTIR of the dual cavity modes generated (1970 cm$^{-1}$ and 2000 cm$^{-1}$). (e) Energy diagram of polariton modes formed by the coupling of $W(CO)_6$ with the coupled-cavity. (f-i) Experimental and simulated linear IR of polaritons. (f and g) Polariton tranmission spectra in a regular cavity detuned to match the resonance of cavity A and B, which can be well fitted by model 1. (h) Experimental polariton tranmission spectra of the coupled cavity and its comparison to the spectra of panels f and g summed together. Clearly, the transmission spectra of the coupled cavity cannot be accounted for by the summed spectra of the two individual cavities. (i) Experimental and simulated linear IR of polaritons in the coupled cavity. Using model 2 (desribed in (b), see main text and S2.2 for details), the experimental spectra can be well simulated. From the simulation result, we also see additional minor polariton peaks appear due to delocalization to the neighboring modes. The new coupling energy scheme is summarized in (e).*

We then examine nonlinear couplings in the coupled-cavity polaritons, by conducting 2D IR spectroscopy(8, 9, 24, 33–38). 2D IR measures the third-order nonlinear response of the systems of interests. The pulse sequence described in Figure 1a excites two vibrational coherences at various time incidences and, therefore, can track the interaction and dynamics of quantum states of the systems (see Methods). For example, when two modes are coupled to each other, e.g. through mechanical or electrostatic coupling, cross peaks appear at the corner defined by the resonance frequencies of the two coupled modes. Thus, it is an ideal way to probe nonlinear couplings between polaritons.

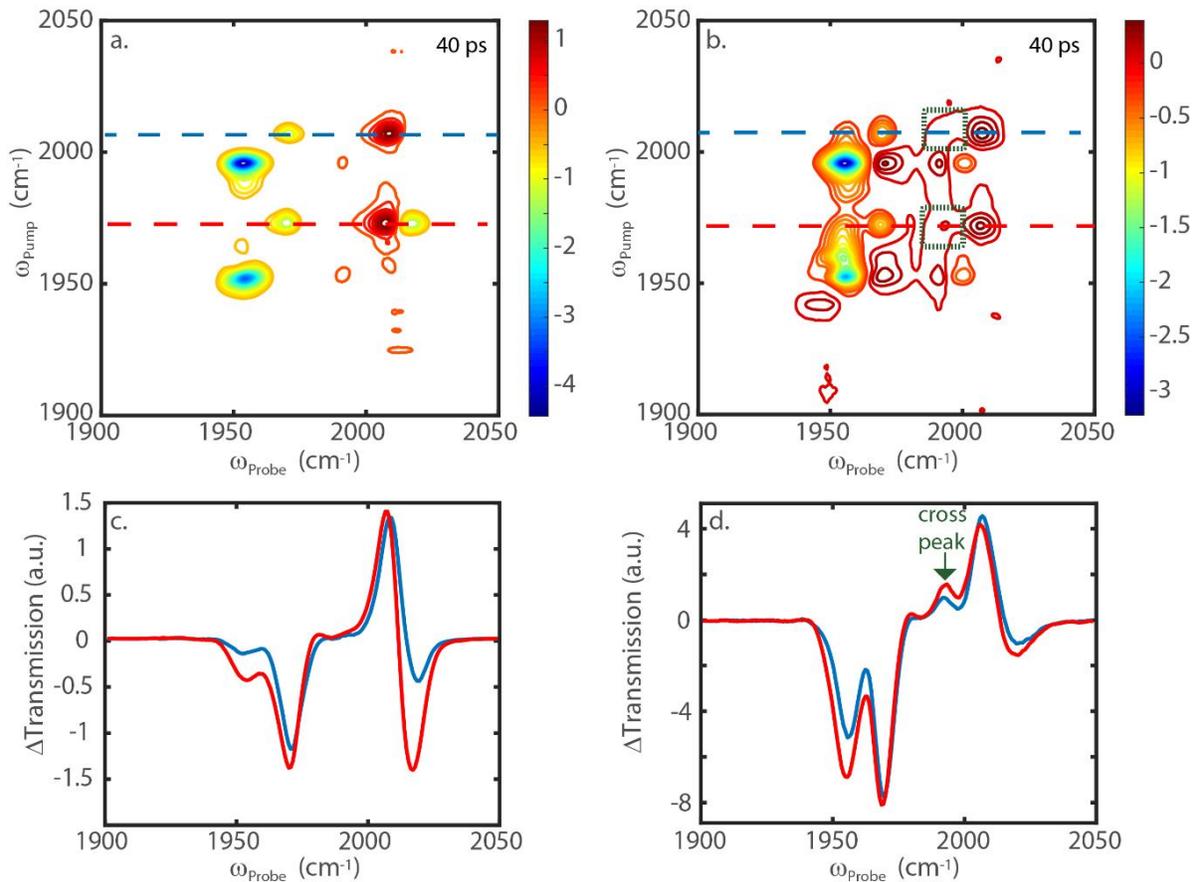

*Figure 2.* 2D IR spectra to show intercavity nonlinear interactions between polaritons. a) 2D IR spectrum of W(CO)$_6$/hexane in two uncoupled cavities (regular cavity A+B). c) Pump spectral cuts of 2D IR in a) at $\omega_{pump} = \omega_{UP1}$ (blue) and $\omega_{LP1}$ (red). b) 2D IR of W(CO)$_6$/hexane in coupled dual cavity. Cross peaks between polaritons from different cavities are observed (in boxed area). d) Pump spectral cuts of 2D IR in b) at $\omega_{pump} = \omega_{UP1}$ (blue) and $\omega_{LP1}$ (red) also show cross peak features.

2D IR spectra of the coupled-cavity polaritons show clear cross peaks ($t_2$ = 40 ps, labeled by dotted green squares in Figure 2b), indicating that polaritons in adjacent cavities couple to each other in the weak-coupling regime. For example, the peak at $\omega_{pump} = \omega_{UP1} = 2010$ and $\omega_{probe} = \omega_{UP2} = 1995$ cm$^{-1}$ suggests by pumping at UP1, the polariton transition at UP2 is perturbed. To ensure the cross peaks are originated from coupled-cavities coupling, we performed 2D IR measurements of the two regular FP cavities, detuned to match the resonance of cavities A and B, separately, and combined their spectra together to simulate polaritons in two uncoupled cavities (Figure 2a and c). We find no cross peaks in this scenario, indicating a lack of intercavity couplings between polaritons.

To quantify the coupling, we take a spectral cut of the 2D spectra at specific $\omega_{pump}$, corresponding to pump-probe spectra by exciting specific quantum states at $\omega_{pump}$, and simulate them. In previous publications by us(7, 24, 33) and Dunkelberger, Simpkins, Owrutsky and co-workers(18, 26), the origin of nonlinear signals of polaritons at long time delay is shown to be a non-equilibrium population of dark states in the cavities. After initial polariton excitations, polariton populations transfer to excited states of dark modes, where the depopulation of ground state leads to the derivative signals near UP due to Rabi splitting contraction, and the first excited state population of dark modes leads to overtone absorptions for the strong absorptive feature at the LP side (S3.1)(7, 24, 26). Similar origins lead to the 2D IR signal of current systems and can be simulated based on model 2 (see Methods and S2.3). The simulated spectra match with the experimental results reasonably well, capturing the derivative features on the UP side as well as the double absorptive features on the LP side (Figure 3a is a spectral cut at $\omega_{pump}= \omega_{LP1}=1970$ cm$^{-1}$; simulations of other spectral cuts are in S3.2). The double absorptive features (at $\omega_{probe} = 1970$ and $= 1955$ cm$^{-1}$) are from 1->2 and 2->3 transition of the dark modes, respectively. Therefore, only the $\omega_{probe} = \omega_{UP2} = 1995$ cm$^{-1}$ peak is purely originated from nonlinear coupling between cavity polaritons (labeled by a green arrow in Figure 2d).

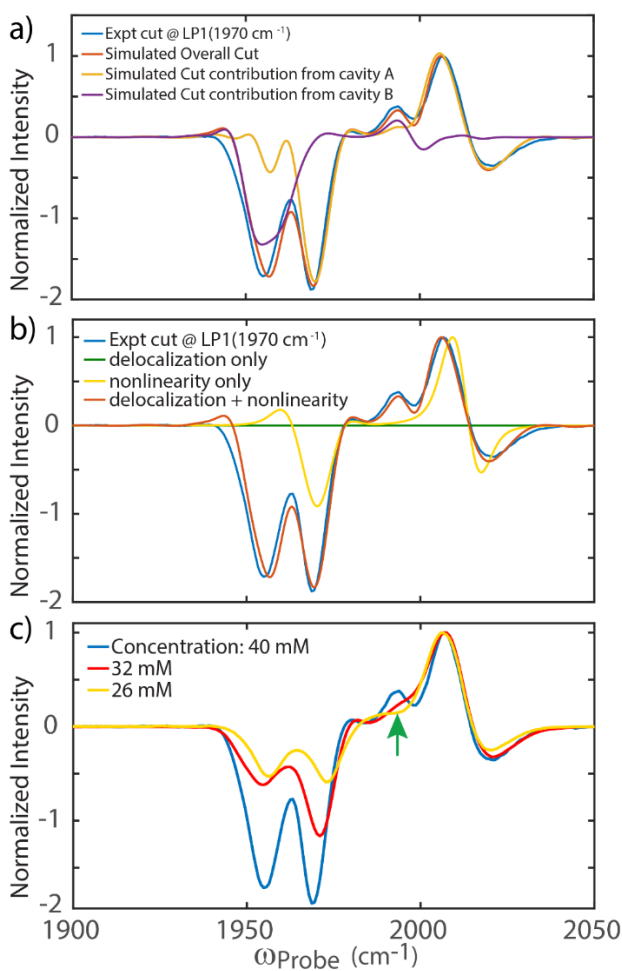

*Figure 3. Comparison between experimental and theoretical nonlinear spectral signals. a) Experimental (blue) and simulated (red) spectral cut at $\omega_{pump}$=1970 cm$^{-1}$ and simulated decomposed contributions from cavity A (yellow) and B (purple); b) Experimental (blue) spectral cut at $\omega_{pump}$=1970 cm$^{-1}$ and simulated cut spectrum with delocalization only (green), molecular nonlinearity only (yellow), and delocalization and molecular nonlinearity together (red); c) Experimental spectral cuts at $\omega_{pump}$=1970 cm$^{-1}$ of different W(CO)$_6$ concentrations (coupling strengths).*

Further physical insights of the intercavity nonlinear signal are obtained from decomposing the simulated 2D spectral into contributions from the cavity A and B paths. The decomposed results show the origin of the cross peak at $\omega_{probe} = \omega_{UP2} = 1995$ cm$^{-1}$ is from nonlinear responses of polaritons of the cavity B path (Figure 3a) : there is an obvious derivative features near $\omega_{probe} = 1995$ cm$^{-1}$ from the nonlinear signal of cavity B (purple trace in Figure 3a), which contributes to the cross peak in the overall cut (red trace in Figure 3a). In contrast, the nonlinear signal of cavity A shows very small peak at $\omega_{probe} = 1995$ cm$^{-1}$ (yellow trace in Figure 3a). Because this spectral cut is obtained by $\omega_{pump} = \omega_{LP1}$ of cavity A, it confirms that by exciting cavity A, polaritons in cavity B can be affected – a real nonlinear coupling between cavities. The mechanism of this nonlinear coupling is that by exciting LP1, dark modes in cavity A are populated to excited states, a portion of which resides in the overlapping volume between cavities A and B, creating nonequilibrium populations that perturb cavity B, and thereby a nonlinear response of polaritons in it.

Further analysis indicates that both cavity delocalization and molecular nonlinearity are key factors in generating intercavity nonlinear interactions between polaritons. To see this, we turn off either cavity delocalization or molecular nonlinearity (Figure 3b). When delocalization is turned off, a spectrum similar to single cavity polaritons is created (yellow trace in Figure 3b), and when molecular nonlinearity is turned off by setting anharmonicity to zero, there are simply no signals (green trace in Figure 3b), because 2D IR measures only nonlinear optical responses of molecules with anharmonicities.

Lastly, we investigate how the intercavity nonlinear interaction between polaritons can be manipulated and controlled by molecular concentration and therefore Rabi splitting: The cross peak becomes stronger as concentration is increased (Figure 3c). This result indicates that there are stronger nonequilibrium populations in the shared mode volume between A and B when polaritons from cavity A are excited. This trend is supported by a simple coupled oscillator model (S3.4) in which each cavity strongly couples to the vibrational modes in its own volume and weakly couples to the ones from the adjacent cavity. From this model, we find that as molecular concentration is increased, the polariton modes contain more molecular wave functions from the adjacent cavity, and, therefore, the two cavity modes share more anharmonic molecular modes (see S3.4), leading to stronger intercavity nonlinear coupling.

The intercavity nonlinear coupling between polaritons is exclusively a result of the joint merits of delocalization of cavity modes and molecular nonlinearity. By combining them, a unique new property is created that does not exist in either mode. This demonstration adds spatial dimension into the parameter space for polariton-based nonlinear optics, and could push IR polaritons into photonic applications on a "2D chip," in IR-based chemical sensing and photonic circuitry. Because the coupled-cavity polaritons contain both strong couplings between molecules and cavities, and weak intercavity nonlinear couplings, coupled-cavity polaritons could be developed into platforms for simulating complex quantum systems. Furthermore, this experiment demonstrates that different cavities can interact with each other through molecules, so it lays a foundation for remote chemistry(4) – enabling chemistry in one cavity, but manipulating other molecules (e.g. as catalysts) in the other cavity.

**Methods**

**Fabrication of couple-cavity optical mirror**

In order to generate two cavity modes with specific frequency difference, two different path lengths need to be achieved within one pair of cavity mirrors. A checkerboard pattern is designed and fabricated on the CaF$_2$ window using photolithography, followed by sputtering deposition of a layer of ZnO and life-off of ZnO deposited on photoresist, thus leaving behind a checkerboard patterned layer of ZnO on $CaF_2$. Dielectric coatings (Thin Film Corp.) are deposited on both the flat $CaF_2$ window and the $CaF_2$ window with patterned ZnO layer to obtain ~96% reflectivity at around 5 μm wavelength. In this work, a dielectric-coated flat $CaF_2$ and a dielectric-coated $CaF_2$ with patterned ZnO (~300 nm) are used in tandem to generate dual cavity modes separated by ~30 $cm^{-1}$ at 5 μm. The frequency separation between the two cavity modes can be tuned by varying the thickness of the ZnO layer.

**Sample Preparation**

The $W(CO)_6$ (Sigma-Aldrich)/coupled-cavity system is prepared in an IR spectral cell (Harrick) containing one flat dielectric $CaF_2$ mirror and one checkerboard-patterned dielectric $CaF_2$ mirror, separated by a 12.5 μm Teflon spacer and filled with $W(CO)_6$/hexane solution (~40mM). The regular $W(CO)_6$/cavity system is prepared in the same way in an IR spectral cell with two flat dielectric $CaF_2$ mirrors.

**2D IR spectroscopy**

Two-dimensional infrared (2D IR) spectroscopy(35, 36) is applied to investigate the light-matter interaction of a $W(CO)_6$/microcavity system as described in previous work(39, 40) (detailed 2D IR set up and data acquisition is described in S1). Briefly, a pump-probe geometry is adopted where three IR pulses (Fig. 1a) interact with sample systems. The first IR pump pulse and probe pulse generate two coherent states in the system in $t_1$ and $t_3$, respectively, which will later be Fourier transformed to frequency domain as $\omega_1$ (pump frequency) and $\omega_3$ (probe frequency). The second IR pump pulse puts the system in a population state during $t_2$. All 2D IR spectra in this work is taken at $t_2 = 40$ ps to avoid interference between pump and probe pulses.

**Simulating Nonlinear Spectra**

Two linear spectra are simulated first, one with all population on the ground vibrational states, corresponding to probe spectra with no IR pump, and another with a certain ground state population lifted to the excited vibrational states, simulating the probe spectra after IR excitation. The differences between the two spectra are taken to simulate pump probe spectra.

**Acknowledgments**

This work is supported by NSF CAREER Award DMR1848215. The fabrication of dual-cavity mirrors was performed in part at the San Diego Nanotechnology Infrastructure (SDNI) of UCSD, a member of the National Nanotechnology Coordinated Infrastructure, which is supported by the National Science Foundation (Grant ECCS-1542148). B.X. is supported by a Roger Tsien Fellowship from the Department of Chemistry and Biochemistry at UC San Diego. We thank Dr. Yuen-Zhou's constructive feedback to this work.

**Author Contributions**

W.X. conceived the original idea and developed the theoretical model for linear and nonlinear spectroscopy in this work. J.W. design and fabricated the coupled-cavity mirrors, analyzed the data, and performed simulations. J.W. and B.X. conducted 2D IR experiments together. J.W., B.X., and W.X. wrote the manuscript.

References

1. Ladd TD, et al. (2010) Quantum computers. *Nature* 464(7285):45–53.

2. Lidar DA (2014) Review of Decoherence-Free Subspaces, Noiseless Subsystems, and Dynamical Decoupling. *Quantum Inf Comput Chem*:295–354.

3. Letokhov VS (1998) *Principles of nonlinear optical spectroscopy* doi:10.1070/PU1998v041n05ABEH000400.

4. Ribeiro RF, Martínez-Martínez LA, Du M, Campos-Gonzalez-Angulo J, Yuen-Zhou J (2018) Polariton chemistry: controlling molecular dynamics with optical cavities. *Chem Sci* 9(30):6325–6339.

5. Georgescu IM, Ashhab S, Nori F (2014) Quantum simulation. *Rev Mod Phys* 86(1). doi:10.1103/RevModPhys.86.153.

6. Simpkins BS, et al. (2015) Spanning Strong to Weak Normal Mode Coupling between Vibrational and Fabry−Pérot Cavity Modes through Tuning of Vibrational Absorption Strength. *ACS Photonics* 2:1460.

7. Ribeiro RF, et al. (2018) Theory for Nonlinear Spectroscopy of Vibrational Polaritons. *J Phys Chem Lett* 9(13):3766–3771.

8. Kurochkin D V., Naraharisetty SRG, Rubtsov I V. (2007) A relaxation-assisted 2D IR spectroscopy method. *Proc Natl Acad Sci U S A* 104(36):14209–14214.

9. Xiang B, et al. (2019) Manipulating Optical Nonlinearities of Molecular Polaritons by Delocalization. *Sci Adv* 5(9):eaax5196.

10. Dorfman KE, Mukamel S (2018) Multidimensional photon correlation spectroscopy of cavity polaritons. *Proc Natl Acad Sci U S A* 115(7):1451–1456.

11. Santhosh K, Bitton O, Chuntonov L, Haran G (2016) Vacuum Rabi splitting in a plasmonic cavity at the single quantum emitter limit. *Nat Commun* 7(1):ncomms11823.

12. Yuen-Zhou J, Menon VM (2019) Polariton chemistry: Thinking inside the (photon) box. *Proc Natl Acad Sci U S A* 116(12):5214–5216.

13. Donati S, et al. (2016) Twist of generalized skyrmions and spin vortices in a polariton superfluid. *Proc Natl Acad Sci U S A* 113(52):14926–14931.

14. Bao W, et al. (2019) Observation of Rydberg exciton polaritons and their condensate in a perovskite cavity. *Proc Natl Acad Sci* 116(41):20274–20279.

15. Rivera N, Rosolen G, Joannopoulos JD, Kaminer I, Soljačić M (2017) Making two-photon processes dominate one-photon processes using mid-IR phonon polaritons. *Proc Natl Acad Sci U S A* 114(52):13607–13612.

16. Lin X, et al. (2017) All-angle negative refraction of highly squeezed plasmon and phonon polaritons in graphene-boron nitride heterostructures. *Proc Natl Acad Sci U S A* 114(26):6717–6721.

17. Van Vugt LK, Piccione B, Cho CH, Nukala P, Agarwal R (2011) One-dimensional polaritons with size-tunable and enhanced coupling strengths in semiconductor nanowires. *Proc Natl Acad Sci U S A* 108(25):10050–10055.

18. Dunkelberger AD, Davidson RB, Ahn W, Simpkins BS, Owrutsky JC (2018) Ultrafast Transmission Modulation and Recovery via Vibrational Strong Coupling. *J Phys Chem A* 122(4):965–971.

19. Liu G, Snoke DW, Daley A, Pfeiffer LN, West K (2015) A new type of half-quantum circulation in a macroscopic polariton spinor ring condensate. *Proc Natl Acad Sci U S A* 112(9):2676–2681.

20. Dunkelberger A, et al. (2019) Saturable Absorption in Solution-Phase and Cavity-Coupled Tungsten Hexacarbonyl. *ACS Photonics*:acsphotonics.9b00703.

21. Imran I, Nicolai GE, Stavinski ND, Sparks JR (2019) Tuning Vibrational Strong Coupling with Co-Resonators. *ACS Photonics* 6(10):2405–2412.

22. Stührenberg M, et al. (2018) Strong Light-Matter Coupling between Plasmons in Individual Gold Bi-pyramids and Excitons in Mono- and Multilayer WSe2. *Nano Lett* 18(9):5938–5945.

23. Flick J, Welakuh DM, Ruggenthaler M, Appel H, Rubio A (2019) Light-Matter Response in Non-Relativistic Quantum Electrodynamics. *ACS Photonics*:acsphotonics.9b00768.

24. Xiang B, et al. (2018) Two-dimensional infrared spectroscopy of vibrational polaritons. *Proc Natl Acad Sci U S A* 115(19):4845–4850.

25. Shalabney A, et al. (2015) Coherent coupling of molecular resonators with a microcavity mode. *Nat Commun* 6(1):5981.

26. Dunkelberger AD, Spann BT, Fears KP, Simpkins BS, Owrutsky JC (2016) Modified relaxation dynamics and coherent energy exchange in coupled vibration-cavity polaritons. *Nat Commun* 7(1):13504.

27. Long JP, Simpkins BS (2015) Coherent coupling between a molecular vibration and fabry-perot optical cavity to give hybridized states in the strong coupling limit. *ACS Photonics* 2(1):130–136.

28. Heylman KD, Knapper KA, Goldsmith RH (2014) Photothermal microscopy of nonluminescent single particles enabled by optical microresonators. *J Phys Chem Lett* 5(11):1917–1923.

29. Liu X, et al. (2014) Strong light-matter coupling in two-dimensional atomic crystals. *Nat Photonics* 9(1):30–34.

30. Neuman T, Esteban R, Casanova D, García-Vidal FJ, Aizpurua J (2018) Coupling of Molecular Emitters and Plasmonic Cavities beyond the Point-Dipole Approximation. *Nano Lett* 18(4):2358–2364.

31. Hartmann MJ, Brandão FGSL, Plenio MB (2006) Strongly interacting polaritons in coupled arrays of cavities. *Nat Phys* 2(12):849–855.

32. Khitrova G, Gibbs HM, Jahnke F, Kira M, Koch SW (1999) Nonlinear optics of normal-mode-coupling semiconductor microcavities. *Rev Mod Phys* 71(5):1591–1639.

33. Xiang B, et al. (2019) State-Selective Polariton to Dark State Relaxation Dynamics. *J Phys Chem A* 123(28):5918–5927.

34. Saurabh P, Mukamel S (2016) Two-dimensional infrared spectroscopy of vibrational polaritons of molecules in an optical cavity. *J Chem Phys* 144(12):124115.

35. Buchanan LE, Xiong W (2018) Two-Dimensional Infrared (2D IR) Spectroscopy. *Encyclopedia of Modern Optics*, pp 164–183.

36. Hamm P, Zanni M (2011) *Concepts and Methods of 2D Infrared Spectroscopy* (Cambridge University Press, Cambridge) doi:DOI: 10.1017/CBO9780511675935.

37. Zhang Z, et al. (2019) Polariton-Assisted Cooperativity of Molecules in Microcavities Monitored by Two-Dimensional Infrared Spectroscopy. *J Phys Chem Lett* 10(15):4448–4454.

38. Fournier JA, Carpenter W, De Marco L, Tokmakoff A (2016) Interplay of Ion-Water and Water-Water Interactions within the Hydration Shells of Nitrate and Carbonate Directly Probed with 2D IR Spectroscopy. *J Am Chem Soc* 138(30):9634–9645.

39. Porter TM, et al. (2019) Direct observation of the intermediate in an ultrafast isomerization. *Chem Sci* 10(1):113–117.

40. Wang J, et al. (2015) Short-Range Catalyst-Surface Interactions Revealed by Heterodyne Two-Dimensional Sum Frequency Generation Spectroscopy. *J Phys Chem Lett* 6(21):4204–4209.

**Supplementary Information**

**S1 2D IR Spectroscopy of W(CO)$_6$ Coupled to Dual Cavity Modes.**

**S2 Theory for Intercavity Polariton-Polariton Interaction**

**S2.1 Model for Transmission of a Single Fabry-Pérot Cavity - Model 1**
**S2.2 Model for the Transmission of Dual Cavity - Model 2**
**S2.3 Pump-Probe Fitting Model Based on Model 2**

**S3 Supplementary Data**

**S3.1 Transient Pump-Probe and 2D IR Spectra of Molecular Vibrational Polariton Systems**
**S3.2 Simulation of Spectral Cuts of 2D IR of 40-mM W(CO)$_6$ in Hexane**
**S3.3 2D IR Spectra of Various Concentrations**
**S3.4 Contributing Components for Polaritons Depending on Coupling Strength (Concentration of W(CO)$_6$ in Hexane)**

# S1 2D IR Spectrometer for Microcavity Systems.

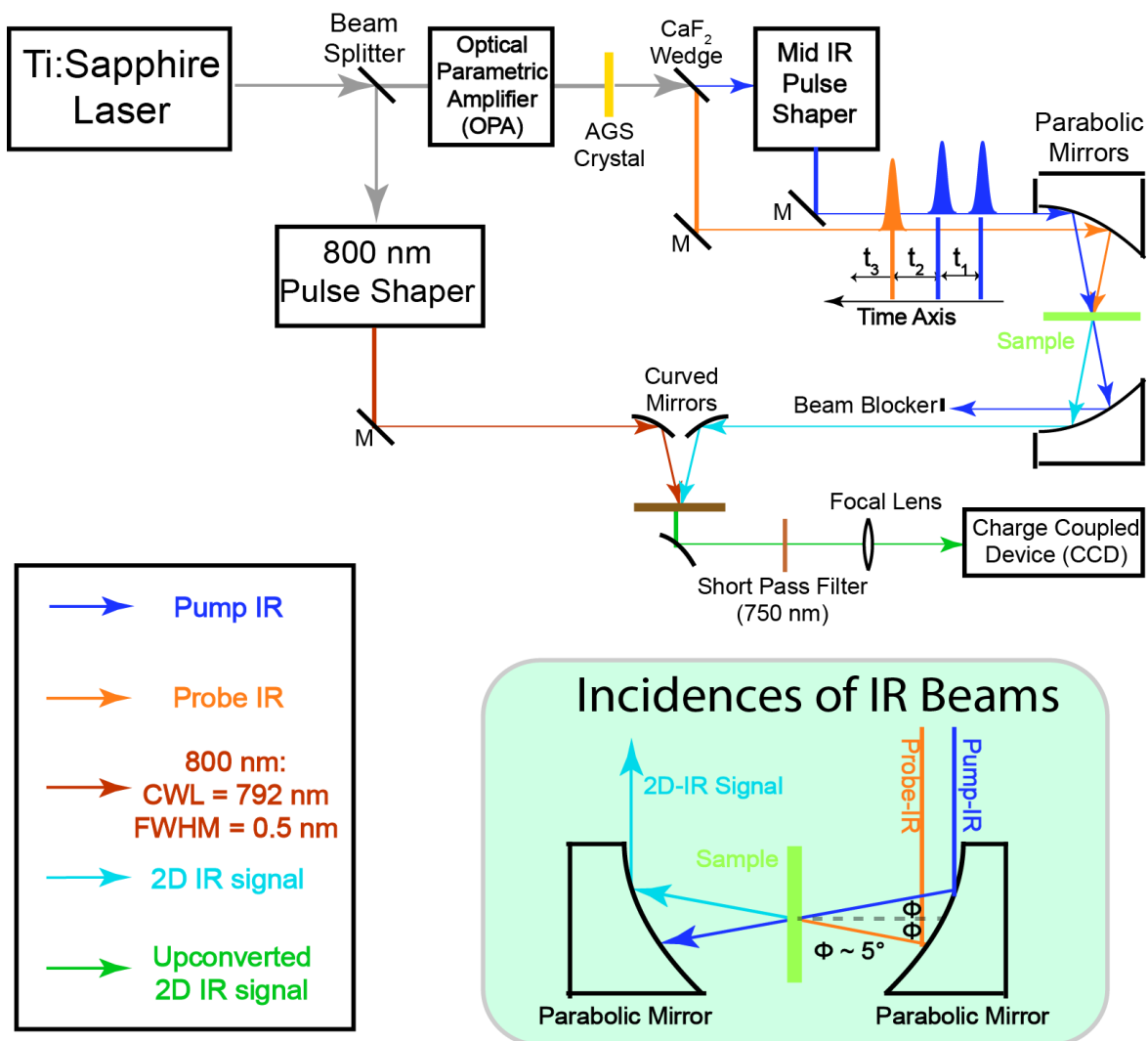

*Figure S1.* Scheme of two-dimensional infrared experimental setup, where the inset shows the incidence of pump and probe IR beams.

Two-dimensional infrared (2D IR) spectroscopy[1–4] is applied to investigate the light-matter interaction of a $W(CO)_6$/microcavity system. The setup scheme is shown in Fig. S1. 800-nm laser pulses (~35 fs, ~5 W, 1 kHz) generated by an ultrafast Ti:Sapphire regenerative amplifier (Astrella, Coherent) are sent into an optical parametric amplifier (OPA) (TOPAS, LightConversion) which outputs tunable near-IR pulses. The near-IR pulses are converted to mid-IR pulses through a difference frequency generation (DFG) process by a type II $AgGaS_2$ crystal (Eksma). After DFG, a $CaF_2$ wedge splits the mid-IR pulse into two parts: the 95% transmitted part is sent into a Ge-Acoustic Optical Modulator based mid IR pulse shaper (QuickShape, PhaseTech) and is shaped to double pulses, which forms the pump beam arm; the 5% reflected is the probe beam. Both pump (~ 1.1 µJ) and probe (~ 0.2 µJ) are focused by a parabolic mirror (f = 10 cm) and overlap spatially at the sample. The output signal is collimated by another parabolic mirror (f = 10 cm) at a symmetric position and is upconverted to an 800-nm beam at a 5%Mg: $LiNbO_3$ crystal. The 800-nm beam that comes out of the OPA passes through an 800-nm pulse shaper which narrows its spectrum in the frequency domain (center wavelength of 791 nm and a FWHM of 0.5 nm or 9.5 cm$^{-1}$).

The pulse sequence is shown in Fig. S1. Two pump pulses and a probe pulse (pulse duration of 100~150 fs) interact with samples at delayed times ($t_1$, $t_2$ and $t_3$). After the first IR pulse, a vibrational coherence is generated, which is converted into a subsequent population state by the second IR pulse and is characterized by scanning $t_1$ (0 to 6000 fs with 20 fs steps) using the mid IR pulse shaper. A rotating frame at $f_0 = 1583$ cm$^{-1}$ is applied to shift the oscillation period to 80 fs and to make the scanning step meet the Nyquist frequency requirement. After waiting for $t_2$, the third IR pulse (probe) is impinged on the sample, and the resulting macroscopic polarization emits an IR signal. This IR signal is upconverted by a narrow-band 800 nm beam. The upconversion process covers the $t_3$ time delay and the 800-nm pulse duration (full width at half maximum = 0.5 nm) determines the scanning length of $t_3$. The monochromator and CCD (Andor) experimentally Fourier transform the upconverted signal, thus generating a spectrum along the $\omega_3$ axis. Numerical Fourier transform of the signal along the $t_1$ axis is required to obtain the spectrum along $\omega_1$. The resulting 2D IR spectra are plotted against $\omega_1$ and $\omega_3$. The $t_2$ time delay is scanned by a computerized delay stage which is controlled by home-written LabVIEW programs to characterize the dynamic features of the system. A rotational stage is mounted on the sample stage to choose the tilt angle and, therefore, the wavevector of the driven polaritons.

## S2 Theory

### S2.1 Model for Transmission of a Single Fabry-Pérot Cavity --- Model 1.

The classical equation for the transmission of a Fabry-Pérot cavity. This expression can provide a basis for relating transient spectra to excited and ground state populations[5–7].

$$T_{cav}(\bar{v}) = \frac{T^2 e^{-\alpha L}}{1 + R^2 e^{-2\alpha L} - 2R e^{-\alpha L} \cos(4\pi n L \bar{v} + 2\varphi)}$$
(S1)

This relationship is based on the frequency dependent absorption coefficient ($\alpha$) and refractive index ($n$) of the material within the cavity. T, R, L, and $\varphi$ is the transmission, reflectivity, thickness, and phase shift of the cavity. We obtain $\alpha$ and $n$ by modeling the dielectric function of the cavity load as a sum of Lorentzian oscillators. The real and imaginary components of the dielectric function, $\varepsilon_1$ and $\varepsilon_2$, are defined as a sum of $i$ Lorentzian oscillators according to

$$\varepsilon_1 = n_{bg}^2 + \sum_i \frac{A_i(v_{0i}^2 - v^2)}{(v_{0i}^2 - v^2)^2 + (\Gamma_i v)^2}, \text{ and}$$
(S2)

$$\varepsilon_2 = \sum_i \frac{A_i \Gamma_i v}{(v_{0i}^2 - v^2)^2 + (\Gamma_i v)^2}$$
(S3)

where $n_{bg}$ is the background refractive index, $A_i$ the amplitude, $v_{0i}$ the resonant frequency, and $\Gamma_i$ the full linewidth associated with the $i^{th}$ oscillator. The frequency-dependent refractive index, $n$, and absorption coefficient, $\alpha$, can be formulated as

$$n = \sqrt{\frac{\varepsilon_1 + \sqrt{\varepsilon_1^2 + \varepsilon_2^2}}{2}},$$
(S4)

$$\alpha = 4\pi v k = 4\pi v \sqrt{\frac{-\varepsilon_1 + \sqrt{\varepsilon_1^2 + \varepsilon_2^2}}{2}}.$$
(S5)

Initial values of $A_i$, $v_{0i}$, and $\Gamma_i$ are chosen to be consistent with the optical response of witness samples, *i.e.*, absorbance for the concentration and pathlengths used.

## S2.2 Model for the Transmission of Dual Cavity --- Model 2

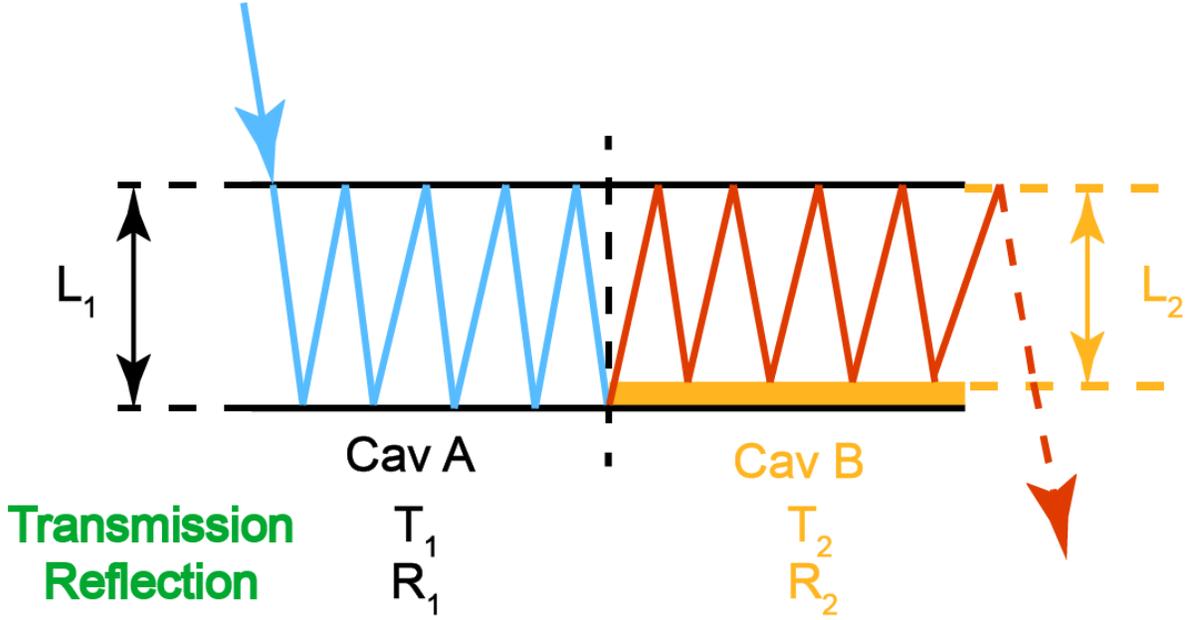

*Figure S2.* Schematic illustration of intercavity hopping, where the key parameters of simulation for both cavities are labeled.

The classical model 2 which involves the intercavity hopping can be derived from the model of FP interferometer[8]. The sum of all the transmitted signals can be expressed as

$$I_{Trans} = \left(\frac{E_2}{E_0}\right)^2. \qquad (S6)$$

$E_2$ is the electric field of all transmitted light and contains the contribution from pure transmission within cavity A area (with only blue paths) and 'hopped' transmission from cavity A to cavity B (with both blue and red paths). More specifically, $E_2(Cav\ A)$ is the electric field of incident light from cavity A area reflected a few times and transmit within A-area; while $E_2(Cav\ A + B)$ is the electric field of incident of incident light from cavity A area reflected a few times in both A-area and B-area (hopped) and transmit in B-area. The two contributions are given

$$\begin{aligned}
E_2(Cav\ A) &= E_0 \Big( t_1 t_1 e^{-\kappa L_1} + t_1 t_1 r_1^2 e^{-3\kappa L_1} e^{i\Delta\phi_1} + t_1 t_1 r_1^4 e^{-5\kappa L_1} e^{2i\Delta\phi_1} + \cdots \\
&\quad + t_1 t_1 r_1^{2(n-1)} e^{-(2n-1)\kappa L_1} e^{(n-1)i\Delta\phi_1} \Big) \\
&= [E_0 t_1 t_1 e^{-\kappa L_1}(1 + \beta + \beta^2 + \beta^3 + \cdots)] \\
&\quad - [E_0 t_1 t_1 r_1^{2n} e^{-(2n+1)\kappa L_1} e^{ni\Delta\phi_1}(1 + \beta + \beta^2 + \beta^3 + \cdots)]
\end{aligned}$$

(S7)

where

$$\beta = r_1^2 e^{-2\kappa L_1} e^{i\Delta\phi_1} \tag{S8}$$

$$(1 + \beta + \beta^2 + \beta^3 + \cdots) = \frac{1}{1-\beta}. \tag{S9}$$

$$E_2(Cav\ A) = \frac{E_0 t_1 t_1 e^{-\kappa L_1}}{1 - r_1^2 e^{-2\kappa L_1} e^{i\Delta\phi_1}} \left(1 - r_1^{2n} e^{-2n\kappa L_1} e^{ni\Delta\phi_1}\right). \tag{S10}$$

$$\begin{aligned}
E_2&(Cav\ A + B) \\
&= E_0 \Big( t_1 t_2 r_1^{2(n-1)} r_2^2 e^{-(2n-1)\kappa L_1} e^{(n-1)i\Delta\phi_1} e^{-2\kappa L_2} e^{i\Delta\phi_2} \\
&\quad + t_1 t_2 r_1^{2(n-1)} r_2^4 e^{-(2n-1)\kappa L_1} e^{(n-1)i\Delta\phi_1} e^{-4\kappa L_2} e^{2i\Delta\phi_2} + \cdots \\
&\quad + t_1 t_2 r_1^{2(n-1)} r_2^{2m} e^{-(2n-1)\kappa L_1} e^{(n-1)i\Delta\phi_1} e^{-2m\kappa L_2} e^{mi\Delta\phi_2} \\
&\quad + \cdots\ (till\ m \to \infty) \Big) \\
&= \left( E_0 t_1 t_2 r_1^{2(n-1)} r_2^2 e^{-(2n-1)\kappa L_1} e^{(n-1)i\Delta\phi_1} e^{-2\kappa L_2} e^{i\Delta\phi_2} \right) (1 + \gamma + \gamma^2 + \gamma^3 + \cdots)
\end{aligned} \tag{S11}$$

where

$$\gamma = r_2^2 e^{-2\kappa L_2} e^{i\Delta\phi_2}. \tag{S12}$$

$$E_2(Cav\ A + B) = \frac{E_0 t_1 t_2 r_1^{2(n-1)} r_2^2 e^{-(2n-1)\kappa L_1} e^{(n-1)i\Delta\phi_1} e^{-2\kappa L_2} e^{i\Delta\phi_2}}{1 - r_2^2 e^{-2\kappa L_2} e^{i\Delta\phi_2}}. \tag{S13}$$

The sum of all the transmitted signals is given by

$$I_{Trans} = \left( \frac{E_2(Cav\ A) + E_2(Cav\ A + B)}{E_0} \right)^2. \tag{S14}$$

Plug in

$$T_i = t_i^2,\ R_i = r_i^2, \alpha = 2\kappa. \qquad (S15)$$

$T_i$, $R_i$, $L_i$ and $\Delta\phi_i$ are the transmission, reflection, cavity longitudinal length and phase shift of corresponding cavity area (A: i = 1; B: i = 2), $\alpha$ is the absorptive coefficient of molecules, and n represents the number of round trips (in A-area) before photon hopping to the adjacent cavity.

The cavity transmission (model 2) can be expressed as

$$T = \left[ \frac{T_1 e^{-\frac{1}{2}\alpha L_1}}{1 - R_1 e^{i\Delta\varphi_1 - \alpha L_1}} \left(1 - R_1^n e^{-n\alpha L_1 + in\Delta\varphi_1}\right) \right.$$

$$\left. + \frac{\sqrt{T_1 T_2} e^{-\frac{1}{2}\alpha L_1}}{1 - R_2 e^{i\Delta\varphi_2 - \alpha L_2}} \left(R_1^{n-1} e^{-(n-1)\alpha L_1 + i(n-1)\Delta\varphi_1} R_2 e^{-\alpha L_2 + i\Delta\varphi_2}\right) \right]^2$$

(S16)

## S2.3 Pump-Probe Fitting Model Based on Model 2.

Based on equation (S16) in S2.2, pump-on ($T_{on}$) and pump-off ($T_{off}$) transmission spectra were simulated separately. By subtracting pump-on with pump-off transmission spectra, pump-probe traces ($T_{pp}$) were obtained. For both pump-on and pump-off spectra, contributions from path A (Fig. 1b, input from cavity A and output from cavity B) and path B (Fig. 1b, input from cavity B and output from cavity A) of incident light. $T_{on}$ has the contributions of path A and B starting from the ground-state concentrations in cavity A, which is proportional to amplitude $A_i$ in $S_2$ and $S_3$ ($\mathcal{H}$-da1-da12) and B ($\mathcal{H}$-da2-da22) after pumping while $T_{off}$ has the ones starting from the ground-state concentrations in cavity A and B (Conc).

For $T_{on}$ and $T_{off}$, the major change is the value of absorption coefficient, $\alpha$. $\alpha(\nu) = 4\pi j(\nu)\nu$, where $j(\nu)$ is the imaginary part of the complex refractive index and $\nu$ is frequency. The following are the quantities which are required for the computation of the transmission with equation S17.

$$j(\nu) = \sqrt{\frac{-\varepsilon_1 + \sqrt{\varepsilon_1^2 + \varepsilon_2^2}}{2}} \tag{S17}$$

where $\varepsilon_1$ and $\varepsilon_2$ are the real and imaginary parts of dielectric constant, expressed as

$$\varepsilon_1 = \varepsilon_{inf} + \sum_{i=1}^{2}\left[\frac{A_i(\nu_i^2 - \nu^2)}{(\nu_i^2 - \nu^2)^2 + (\Gamma_i \nu)^2}\right], \quad \varepsilon_2 = \sum_{i=1}^{2}\left[\frac{A_i \Gamma_i \nu}{(\nu_i^2 - \nu^2)^2 + (\Gamma_i \nu)^2}\right] \tag{S18}$$

where we set the background dielectric constant at infinite frequency to be $\varepsilon_{inf} = 2.0135$, $\nu_i$ are the frequencies of the $0 \to 1$ and $1 \to 2$ asymmetric stretch transitions of W(CO)$_6$ given by $\nu_1 = 1983$ cm$^{-1}$, and $\nu_2 = 1968$ cm$^{-1}$, and the $\Gamma_i$ are the linewidths of the corresponding vibrational modes ($\Gamma_1$ and $\Gamma_2$ are 3.0 and 4.5 cm$^{-1}$, respectively). The oscillator strength $A_i$ at corresponding states (ground, first excited and second excited states) is one of the variables in the fitting that needs to be optimized. For pump-on spectra, $A_1 =$ $\mathcal{H}$-da1-da12 (path A) or $\mathcal{H}$-da2-da22 (path B), $A_2 =$ da1 (path A) or da2 (path B) and $A_3 =$ da12 (path A) or da22 (path B). For pump-off spectra, $A_1 = \mathcal{A}$ (both path A and B), $A_2 = A_3 = 0$ (both path A and B). The number of round trips in the starting cavity (cavity A in path A and cavity B in path B), (n-1), will be $N_1$ (path A) and $N_2$ (path B). The physical meanings of all the fitting parameters are shown below and the values of them will be listed in S3.2.

**Physical Meanings of Fitting Parameters:**

**$L_1$:** thickness of cavity A

**$L_2$:** thickness of cavity B

**$T_1$:** transmission of cavity mirror for cavity A

**$T_2$:** transmission of cavity mirror for cavity B

**$N_1$:** number of round trips in cavity A of path cavity A

**$N_2$:** number of round trips in cavity B of path cavity B

**$\mathcal{A}$:** Amplitude of molecular transitions, related to static concentration of molecules

**da1:** Amplitude of molecular transitions, related to change of ground state concentration of cavity A

**da12:** Amplitude of molecular transitions, related to change of the first excited vibrational state concentration of cavity A

**da2:** Amplitude of molecular transitions, related to change of ground state concentration of cavity B

**da22:** Amplitude of molecular transitions, related to change of the first excited vibrational state concentration of cavity B

# S3 Supplementary Data

**S3.1 Transient Pump-Probe and 2D IR Spectra of Molecular Vibrational Polariton Systems.**

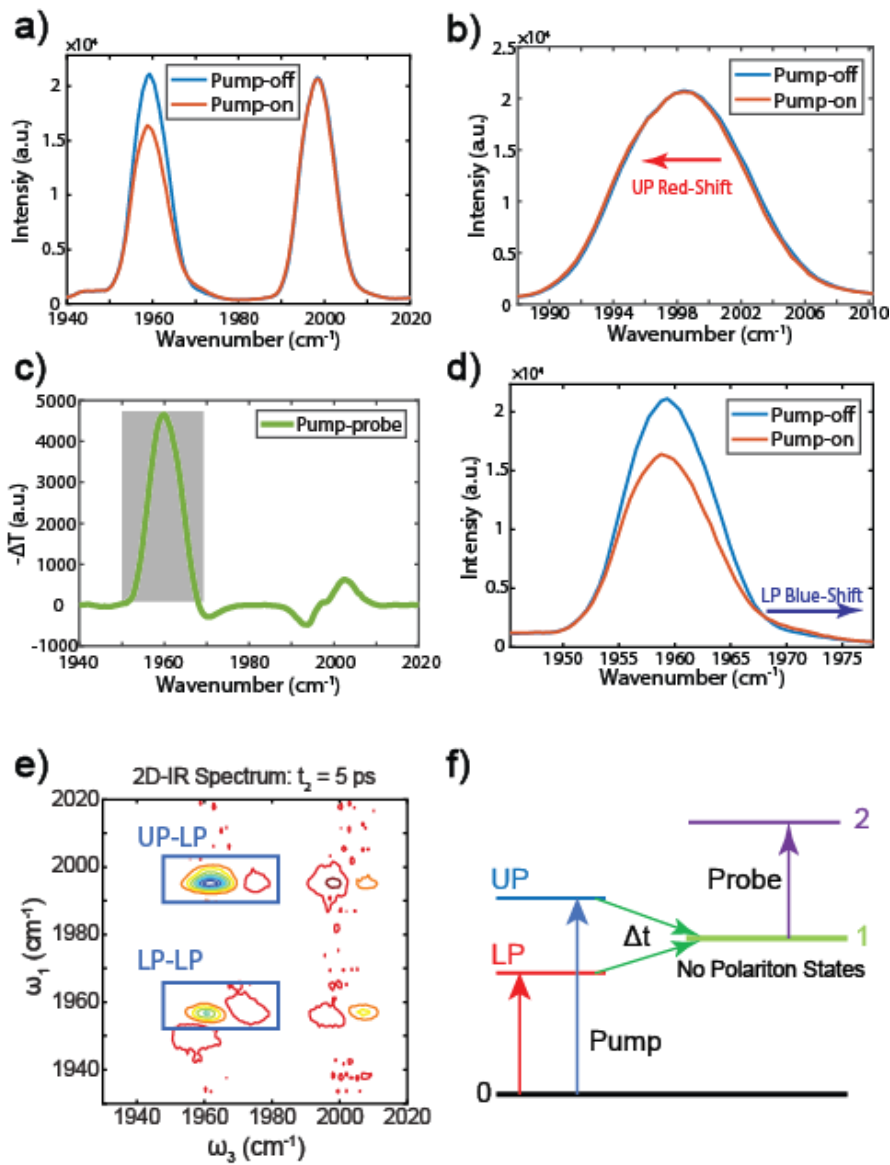

*Figure S3.* Pump probe, 2D IR spectra of molecular polaritons. a) Pump-on and pump-off spectra of strongly coupled $W(CO)_6$/hexane system in transient pump-probe experiment at $t_2$ = 25 ps; b) UP branch zoom-in; c) pump-probe spectrum at $t_2$ = 25 ps; d) LP branch zoom-in; (e) 2D IR spectrum of strongly coupled $W(CO)_6$/hexane system at $t_2$ = 5 ps; (f) Schematic illustration of the population transfer process when the polariton system is in equilibrium ($t_2$ > 3 ps).

Fig. S3 shows representative transient pump-probe spectra[9–11] under strong coupling conditions (1 Vibrational mode + 1 Cavity mode), along with 1D transmission polariton spectra (strongly coupled $W(CO)_6$ in hexane) under pump-on and pump-off conditions, at $t_2$ > 3 ps (Fig. S3a). When the pump is turned on, the UP resonance undergoes a shift towards a lower frequency (Fig. S3b). Under the same condition, the LP lineshape acquires a small positive shoulder appears at higher frequency, which corresponds to a blue shift (Fig. S3d). These shifts are small but consistent and result in a derivative lineshape in the transient pump-probe spectrum (Fig. S3c). The peak-shift is induced by the Rabi splitting contraction which arises due to the pump-induced reduction of molecular ground-state population. The substantially reduced LP transmission upon pumping, and consequently the absorptive lineshape in the pump probe spectra, results from that the dark mode overtone $v_{12}$ transition (from first excited to second excited states, purple arrow in Fig. S3f) is near resonance with LP transition. As a result, $v_{12}$ become visible through the LP transmission window. Thus, when LP and $v_{12}$ are near resonance, the appearance of strong absorptive transient signal at $\omega_{LP}$ is a signature of populating first excited state of dark modes.

While pump probe spectroscopy allows following polariton to dark state dynamics, state-selective 2D IR spectrum (Fig. S3e) enables disentangle the dynamics: The UP-LP peak labeled in Fig. S3e (left-top) represents the population transfer from UP state to dark mode while the LP-LP peak (Fig. S3e, left-bottom) is mainly due to the LP to dark mode population transfer. As summarized in Fig. S3f, it is believed that the UP/LP population transfer to dark modes in a fast timescale, which subsequently makes the dark mode $v_{12}$ appear in the pump probe or 2D spectra. We can learn the polariton dynamics by measuring dynamics of LP peak in pump probe spectra (integrating over the transient pump probe peak near LP position, e.g. shaded area in Fig.S3c)), and dynamics of UP-LP and LP-LP peaks from 2D IR spectra (integrating 2D spectral peaks at UP-LP and LP-LP area, e.g, green boxes in Fig. S3e).

## S3.2 Simulation of Spectral Cuts of 2D IR of 40-mM W(CO)$_6$ in Hexane

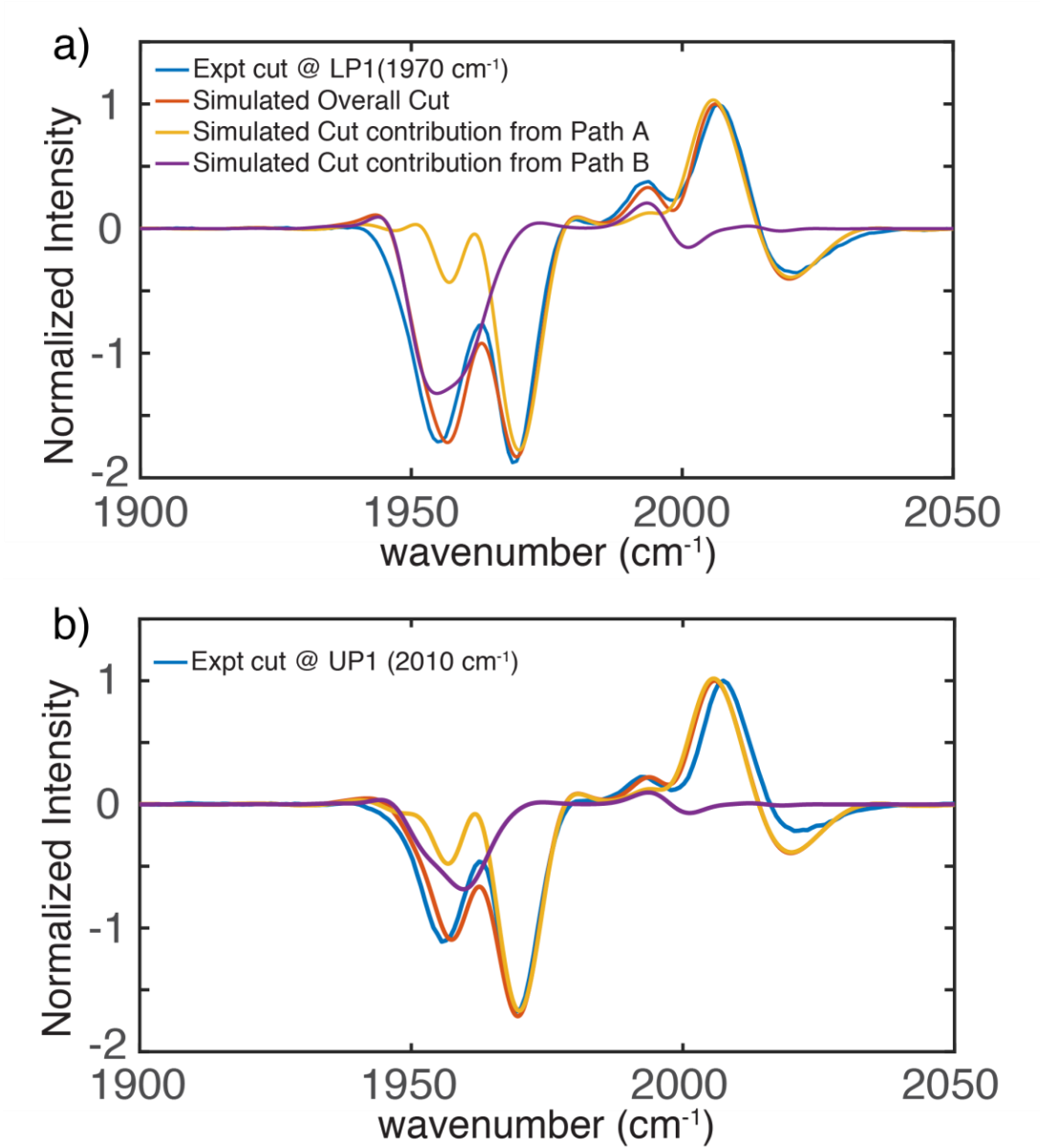

*Figure S4. a)* 2D IR spectral cut at $\omega_1 = \omega_{LP1}$ (top panel) and *b)* 2D IR spectral cut at $\omega_1 = \omega_{UP1}$ (bottom panel), with simulation results of intercavity hopping. In both panels, the red trace is the overall results, while the orange and purple traces represent the contribution from path A and B shown in Fig. 1b.

Fig. S 4a) and b show the experimental data and simulation results of LP1 and UP1 spectral cuts. The detailed interpretation of LP1 spectral cuts has been made in the main manuscript when describing Fig. 3a, while the results of UP1 spectral cuts show similar features. When UP1 from cavity A is pumped, UP2

from cavity B is perturbed, which results in the UP1-UP2 cross-peak (shown in Fig. 2b, the top cross-peak in boxed area). The fitting parameters of both spectral cuts and individual components are shown as follow

|  | $\mathcal{H}$ (cm$^{-1}$) | da1(cm$^{-1}$) | da12(cm$^{-1}$) | da2(cm$^{-1}$) | da22(cm$^{-1}$) |
|---|---|---|---|---|---|
| Cut at LP1 (1970 cm$^{-1}$) | 2600 | 1230 | 10 | 230 | 50 |
| Cut at UP1 (2010 cm$^{-1}$) | 2600 | 1230 | 50 | 130 | 15 |

$L_1$=0.002290; $L_2$=0.002254; $T_1$=0.11; $T_2$=0.065; $N_1$=7; $N_2$=10 were used for both of the spectral cuts at LP1 and UP1.

**S3.3 2D IR Spectra of Various Concentrations.**

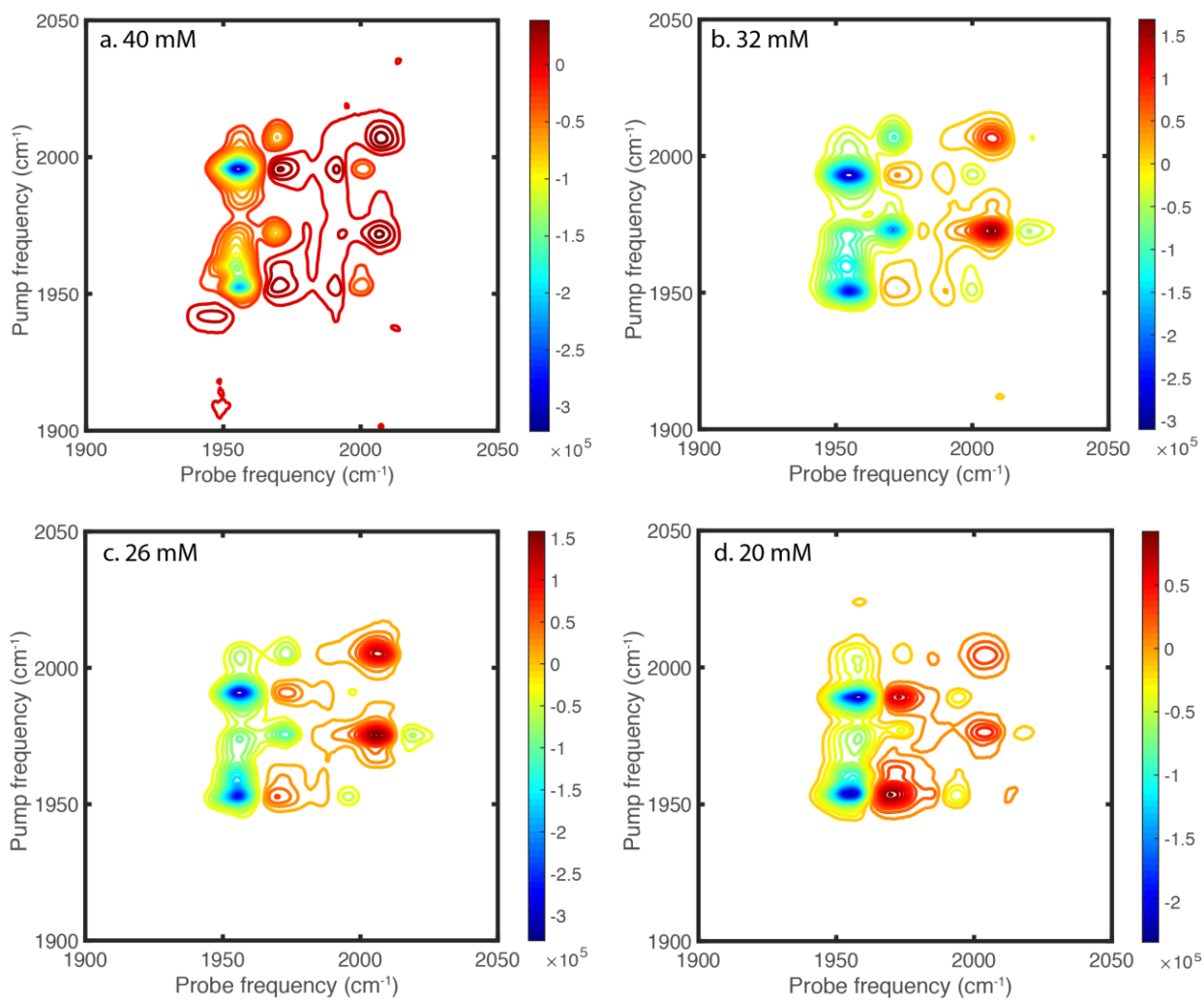

***Figure S5.*** *2D IR spectra of W(CO)$_6$/hexane in dual-cavity at t$_2$ = 40 ps with molecular concentrations of (a) 40 mM, (b) 32 mM, (c) 26 mM, and (d) 20 mM.*

**S3.4 Contributing Components for Polaritons Depending on Coupling Strength (Concentration of W(CO)$_6$ in Hexane).**

The eigen states (four polaritonic states: LP1, UP1, LP2 and UP2) can be derived by the following equation (S19), where the matrix represents the coupling between four original states. Both Vib A and Vib B are the molecular modes and they would strongly couple to cavity A and B modes separately with the same coupling strength, G. Meanwhile, there is weak coupling happening (with coupling strength, g) between Vib A and cavity B as well as Vib B and cavity A. By changing the values of G and g (tuned by molecular concentration, the ratio between G and g are kept same), the contributions of each components (Hopfield coefficients) to LP1 and UP1 states can be obtained accordingly and are shown in Table S1~S3 with different combinations of G and g. It is found that as concentration, and therefore G and g become larger, the more Vib B and Cavity B contribute to the formation of UP1 and LP1.

$$det\begin{pmatrix} Vib\ A & 0 & G & g \\ 0 & Vib\ B & g & G \\ G & g & Cavity\ A & 0 \\ g & G & 0 & Cavity\ B \end{pmatrix} = 0 \qquad S(19)$$

**Table S1.** *Contributing components for UP1 and LP1 with G=20cm$^{-1}$, g=5 cm$^{-1}$*

|     | Vib A (1983 cm$^{-1}$) | Vib B (1983 cm$^{-1}$) | Cavity A (2000 cm$^{-1}$) | Cavity B (1970 cm$^{-1}$) |
| --- | --- | --- | --- | --- |
| UP1 | 0.282 | 0.052 | 0.640 | 0.026 |
| LP1 | 0.635 | 0.034 | 0.275 | 0.057 |

**Table S2.** *Contributing components for each polariton with G=18cm$^{-1}$, g=4.5 cm$^{-1}$*

|     | Vib A (1983 cm$^{-1}$) | Vib B (1983 cm$^{-1}$) | Cavity A (2000 cm$^{-1}$) | Cavity B (1970 cm$^{-1}$) |
| --- | --- | --- | --- | --- |
| UP1 | 0.268 | 0.045 | 0.666 | 0.021 |
| LP1 | 0.661 | 0.028 | 0.262 | 0.049 |

**Table S3.** *Contributing components for each polariton with G=16cm$^{-1}$, g=4 cm$^{-1}$*

|     | Vib A (1983 cm$^{-1}$) | Vib B (1983 cm$^{-1}$) | Cavity A (2000 cm$^{-1}$) | Cavity B (1970 cm$^{-1}$) |
| --- | --- | --- | --- | --- |
| UP1 | 0.252 | 0.038 | 0.694 | 0.016 |
| LP1 | 0.689 | 0.023 | 0.246 | 0.042 |

References

1. Hamm, P. & Zanni, M. *Concepts and Methods of 2D Infrared Spectroscopy*. (Cambridge University Press, 2011).

2. Saurabh, P. & Mukamel, S. Two-dimensional infrared spectroscopy of vibrational polaritons of molecules in an optical cavity. *J. Chem. Phys.* **144**, 124115 (2016).

3. Mukamel, S. *Principles of nonlinear optical spectroscopy.* (New York: Oxford university press, 1995).

4. Buchanan, L. E. & Xiong, W. Two-Dimensional Infrared (2D IR) Spectroscopy. in *Encyclopedia of Modern Optics* **2**, 164–183 (2018).

5. Dunkelberger, A. D., Spann, B. T., Fears, K. P., Simpkins, B. S. & Owrutsky, J. C. Modified relaxation dynamics and coherent energy exchange in coupled vibration-cavity polaritons. *Nat. Commun.* **7**, 13504 (2016).

6. Long, J. P. & Simpkins, B. S. Coherent coupling between a molecular vibration and fabry-perot optical cavity to give hybridized states in the strong coupling limit. *ACS Photonics* **2**, 130–136 (2015).

7. Khitrova, G., Gibbs, H. M., Jahnke, F., Kira, M. & Koch, S. W. Nonlinear optics of normal-mode-coupling semiconductor microcavities. *Rev. Mod. Phys.* **71**, 1591–1639 (1999).

8. Heald, Mark A., Marion, J. B. *Classical Electromagnetic Radiation*. (Saunders College Publication, 1995).

9. Ribeiro, R. F. *et al.* Theory for Nonlinear Spectroscopy of Vibrational Polaritons. *J. Phys. Chem. Lett.* **9**, 3766–3771 (2018).

10. Xiang, B. *et al.* Two-dimensional infrared spectroscopy of vibrational polaritons. *Proc. Natl. Acad. Sci.* **115**, 4845–4850 (2018).

11. Xiang, B. *et al.* State-Selective Polariton to Dark State Relaxation Dynamics. *J. Phys. Chem. A* **123**, 5918–5927 (2019).